\documentclass[12pt]{article}
\usepackage{graphics} 
\usepackage{cite}
\usepackage{latexsym}
\newcommand   {\etal}    {{\it et~al.}}
\newcommand  {\APS}      {{\it Adv. Polym. Sci.\ }}

\newcommand  {\EBJ}      {{\it Eur.\ Biophys.\ J.\ }}
\newcommand  {\EL}       {{\it Europhys.\ Lett.\ }}

\newcommand  {\JCC}      {{\it J.\ Comput.\ Chem.\ \ }}

\newcommand  {\JCP}      {{\it J.\ Chem.\ Phys.\ }}

\newcommand  {\JPCM}     {{\it J.\ Phys.: Condens. Matter\ }}

\newcommand  {\JSP}      {{\it J.\ Stat.\ Phys.\ }}

\newcommand  {\Mac}      {{\it Macromolecules\ }}

\newcommand  {\MP}       {{\it Molec.\ Phys.\ }}

\newcommand  {\PNAS}     {{\it Proc.\ Natl.\ Acad.\ Sci.\ USA\ }}

\newcommand  {\PRL}      {{\it Phys.\ Rev.\ Lett.\ }}

\newcommand  {\TRA}      {{\it IEEE Trans. Rob. Autom.\ }}

\newcommand{\beq}{\begin{equation}}
\newcommand{\eeq}{\end{equation}}
\newcommand{\beqa}{\begin{eqnarray}}
\newcommand{\eeqa}{\end{eqnarray}}
\newcommand{\bea}{\begin{eqnarray}}
\newcommand{\eea}{\end{eqnarray}}
\newcommand {\fig}[3]{\resizebox{#2}{!}{\rotatebox{#3}{\includegraphics{#1}}}}

\newcommand   {\ev}[1]   {\langle #1\rangle}
\newcommand   {\Ca}      {C${}_{\alpha}$}
\newcommand   {\Cb}      {C${}_{\beta}$}
\newcommand   {\Cp}      {C${}^{\prime}$}
\newcommand   {\Eloc}    {E_{\mbox{{\scriptsize loc}}}}

\newcommand   {\Esa}     {E_{\mbox{{\scriptsize sa}}}}
\newcommand   {\Ehb}     {E_{\mbox{{\scriptsize hb}}}}
\newcommand   {\Eaa}     {E_{\mbox{{\scriptsize AA}}}}
\newcommand   {\eloc}    {\epsilon_{\mbox{{\scriptsize loc}}}}
\newcommand   {\esa}     {\epsilon_{\mbox{{\scriptsize sa}}}}
\newcommand   {\ehb}     {\epsilon_{\mbox{{\scriptsize hb}}}}
\newcommand   {\eaa}     {\epsilon_{\mbox{{\scriptsize AA}}}}
\newcommand   {\shb}     {\sigma_{\mbox{{\scriptsize hb}}}}
\newcommand   {\saa}     {\sigma_{\mbox{{\scriptsize AA}}}}

\newcommand   {\dpv}     {\delta{\bar \phi}}
\newcommand   {\pv}      {{\bar \phi}}
\newcommand   {\psv}     {{\bar \psi}}
\newcommand   {\rv}      {{\bar r}}

\newcommand   {\Av}      {{\bf A}}
\newcommand   {\Gv}      {{\bf G}}
\newcommand   {\unit}    {{\bf 1}}
\newcommand   {\Lv}      {{\bf L}}
\newcommand   {\Pacc}    {P_{\mbox{{\scriptsize acc}}}}

\begin{document}

\begin{flushright}
LU TP 00-55\\
Revised version\\
March 28, 2001
\end{flushright}

\vspace{0.4in}

\begin{center}

{\LARGE \bf Monte Carlo Update for Chain Molecules:} 

{\LARGE \bf Biased Gaussian Steps in Torsional Space} 

\vspace{.3in}

\large
Giorgio Favrin, Anders Irb\"ack and Fredrik 
Sjunnesson\footnote{E-mail: favrin,\,anders,\,fredriks@thep.lu.se}\\   
\vspace{0.10in}
Complex Systems Division, Department of Theoretical Physics\\ 
Lund University,  S\"olvegatan 14A,  S-223 62 Lund, Sweden \\
{\tt http://www.thep.lu.se/complex/}\\

\vspace{0.3in}	

Submitted to \JCP

\end{center}
\vspace{0.3in}
\normalsize
Abstract:\\
We develop a new elementary move for simulations of polymer chains  
in torsion angle space. The method is flexible and easy to implement. 
Tentative updates are drawn from a
(conformation-dependent) Gaussian distribution that favors
approximately local deformations of the chain. The degree
of bias is controlled by a parameter $b$. The method is tested 
on a reduced model protein with 54 amino acids and the Ramachandran 
torsion angles as its only degrees of freedom, for different $b$. 
Without excessive fine tuning, we find that the effective step size can be 
increased by a factor of three compared to the unbiased $b=0$ case.
The method may be useful for kinetic studies, too. 
 
\newpage

\section{Introduction}

Kinetic simulations of protein folding are notoriously difficult. 
Thermodynamic simulations may use unphysical moves and are therefore
potentially easier, but existing methods need improvement. Three
properties that a successful thermodynamic algorithm must possess are 
as follows. First and foremost, it must be able to alleviate the 
multiple-minima problem. Methods like the multicanonical 
algorithm~\cite{Berg:92,Hansmann:93} and simulated 
tempering~\cite{Lyubartsev:92,Marinari:92,Irback:95} 
try to do so by the use of generalized ensembles. Second, it must 
provide an efficient evolution of large-scale properties of
unfolded chains. The simple pivot method~\cite{Lal:69} does 
remarkably well~\cite{Madras:88} in that respect.  
Third, it must be able to alter local properties of folded chains
without causing too drastic changes in their global structure. This
paper is concerned with the third problem, which is important if 
the backbone potentials are stiff and especially if the   
mobility is restricted to the biologically most relevant torsional
degrees of freedom.    

An update that rearranges a restricted section of the chain 
without affecting the remainder is local. For chains with 
flexible or semiflexible backbones, there exists a 
variety of local updates, ranging from simple single-site 
moves to more elaborate 
methods~\cite{Escobedo:95,Frenkel:96,Vendruscolo:97,Wick:00,Escobedo:00}
where inner sections are removed and then regrown site by site 
in a configurational-bias manner~\cite{Frenkel:92,dePablo:92}. 
However, these methods break down if bond lengths and bond angles 
are completely rigid. 

The problem of generating local deformations of chains with only torsional 
degrees of freedom was analyzed in a classic paper by G\=o and 
Scheraga~\cite{Go:70}. Based on this analysis, Dodd~\etal~\cite{Dodd:93} 
devised the first proper Monte Carlo algorithm of this type, the    
concerted-rotation method. This method works with seven adjacent 
torsion angles along the chain. One of these angles is turned by a 
random amount. Possible values of the remaining six angles are 
then determined by numerically solving a set of equations that
guarantee that the move is local. The new conformation is finally 
drawn from the set of all possible solutions to this so-called
rebridging problem.  
Variations and generalizations of this method have been discussed 
by several groups~\cite{Hoffmann:96,Pant:95,Mavrantzas:99}. There are 
also methods~\cite{Leonitidis:94,Deem:96,Wu:99a,Wu:99b,Uhlherr:00}  
that combine elements of the configurational-bias and concerted-rotation 
approaches. One of these methods~\cite{Wu:99b} 
uses an analytical rebridging scheme, inspired by the 
solution for a similar problem in robotic control~\cite{Manocha:94}.      

The concerted-rotation approach is a powerful method that can
generate large local deformations by
finding the discrete solutions to the rebridging problem. 
However, the method is not easy to implement and large 
local deformations may be difficult to accomplish 
if, for example, the chain is folded and has   
bulky side groups. Hence, there are situations where 
this method is not the obvious choice.  

In this paper, we discuss a different and less sophisticated 
type of Monte Carlo move in torsion angle space. This algorithm
is by nature a ``small-step'' algorithm so large local deformations 
cannot take place. Drastic global changes would still occur if the 
steps were random. To avoid that, a biasing probability is 
introduced. The method becomes approximately local if the bias
is made strong. Compared to a strictly local update, 
this method has the disadvantage that a much smaller 
part of the energy function is left unchanged, so 
the CPU time per update is larger. However, this problem is 
not too severe for moderate chain lengths. Moreover, both our 
method and strictly local ones are typically combined 
with some truly nonlocal update like pivot, and such an 
update is not faster than ours. 
           
The algorithm proceeds as follows. We consider $n$ 
torsion angles $\phi_i$, where $n=8$ in our calculations.  
To update these angles, we introduce a conformation-dependent  
$n\times n$ matrix $\Gv$ such that $\dpv^T\Gv\dpv\approx0$ for changes 
$\dpv=(\delta\phi_1,\ldots,\delta\phi_n)$ that correspond 
to local deformations. The steps $\dpv$ 
are then drawn from the Gaussian distribution
\beq 
P(\dpv)\propto\exp\left[-\frac{a}{2}\dpv^T(\unit+b\Gv)\dpv\right]\,,
\eeq
where $\unit$ denotes the $n\times n$ unit matrix and $a$
and $b$ are tunable parameters. The parameter 
$a$ controls the acceptance rate, whereas $b$
sets the degree of bias. The new conformation is finally subject to 
an accept/reject step. Important to the implementation of  
the algorithm is that the matrix $\Gv$ is non-negative and 
symmetric. Hence, it is possible to take the ``square root'' 
of $\unit+b\Gv$, which facilitates the calculations. 

This method, which is quite general, is tested on a reduced model 
protein~\cite{Irback:00} with 54 amino acids and the Ramachandran 
torsion angles as its only degrees of freedom. This chain forms  
a three-helix bundle in its native state and exhibits an abrupt 
collapse transition that coincides with its folding transition.  
The performance of the method is studied 
both above and below the folding temperature,
for different values of the parameters $a$ and $b$. 
For a suitable choice of $b$, we find that the effective 
step size can be increased by a factor of three in the folded phase, 
compared to the unbiased $b=0$ case. The optimal value of $b$ corresponds 
to a relatively strong bias, that is an approximately local update.  

\section{The Model}
\label{sec:2}
 
In our calculations, we consider a reduced protein 
model~\cite{Irback:00} where each amino acid is represented 
by five or six atoms. The three backbone atoms N, \Ca\ and 
\Cp\ are all included, whereas the side chain is represented
by a single atom, \Cb. The \Cb\ atom can be hydrophobic, polar 
or absent, which means that there are three different types of amino acids
in the model. For a schematic illustration of the chain
representation, see Fig.~\ref{fig:1}.  

All bond lengths, bond angles and peptide 
torsion angles ($180^\circ$) are held fixed, which leaves us 
with two degrees of freedom per amino acid, the 
Ramachandran torsion angles (see Fig.~\ref{fig:1}).

The energy function 
\beq
E=\Eloc+\Esa+\Ehb+\Eaa
\label{e}\eeq
is composed of four terms. 
The local potential $\Eloc$ has a standard form with threefold symmetry,
\beq
\Eloc=\frac{\eloc}{2}\sum_i(1 + \cos3\phi_i)\,.
\eeq
The self-avoidance term $\Esa$ is given by a hard-sphere potential
of the form 
\beq
\Esa=\esa\mathop{{\sum}'}_{i<j}
\bigg(\frac{\sigma_{ij}}{r_{ij}}\bigg)^{12}\,,
\label{sa}\eeq 
where the sum runs over all possible atom pairs except those  
consisting of two hydrophobic \Cb. The hydrogen-bond term $\Ehb$
is given by
\beq
\Ehb= \ehb \sum_{ij}u(r_{ij})v(\alpha_{ij},\beta_{ij})\,,
\label{hb}\eeq
where $i$ and $j$ represent H and O atoms (see Fig.~\ref{fig:1}), 
respectively, and 
\begin{eqnarray} 
u(r_{ij})&=&  5\bigg(\frac{\shb}{r_{ij}}\bigg)^{12} - 
        6\bigg(\frac{\shb}{r_{ij}}\bigg)^{10}\\
v(\alpha_{ij},\beta_{ij})&=&\left\{ 
        \begin{array}{ll}
 \cos^2\alpha_{ij}\cos^2\beta_{ij} & \ \alpha_{ij},\beta_{ij}>90^{\circ}\\
 0                      & \ \mbox{otherwise}
         \end{array} \right. 
\end{eqnarray}
In these equations, $r_{ij}$ denotes the HO distance, $\alpha_{ij}$ the NHO 
angle, and $\beta_{ij}$ the HO\Cp\ angle. Finally, the hydrophobicity 
term $\Eaa$ has the form
\beq
\Eaa=\eaa\sum_{i<j}\bigg[
\bigg(\frac{\saa}{r_{ij}}\bigg)^{12}
-2\bigg(\frac{\saa}{r_{ij}}\bigg)^6\,\bigg]\,,
\eeq
where both $i$ and $j$ represent hydrophobic \Cb. In the following, 
$kT$ is given
in dimensionless units, in which $\ehb=2.8$ and $\eaa=2.2$. Further details of
the model, including numerical values of all the parameters, can be found
in Ref.~\cite{Irback:00}.  

In this model, we study a designed three-helix-bundle protein with
54 amino acids. In Ref.~\cite{Irback:00}, it was demonstrated that this
sequence indeed forms a stable three-helix bundle, except for a
twofold topological degeneracy, and that it has a first-order-like 
folding transition that coincides with the collapse transition. 
It should be noted that
these properties are found without resorting to the widely used but
drastic G\=o approximation~\cite{Go:78}, where interactions
that do not favor the desired structure are ignored.   

\section{The Algorithm}

We now turn to the algorithm, which we describe assuming the particular
chain geometry defined in Sec.~\ref{sec:2}. That this scheme 
can be easily generalized to other types of chains will be evident.     
  
\begin{figure}[t]
 \begin{center}
  \fig{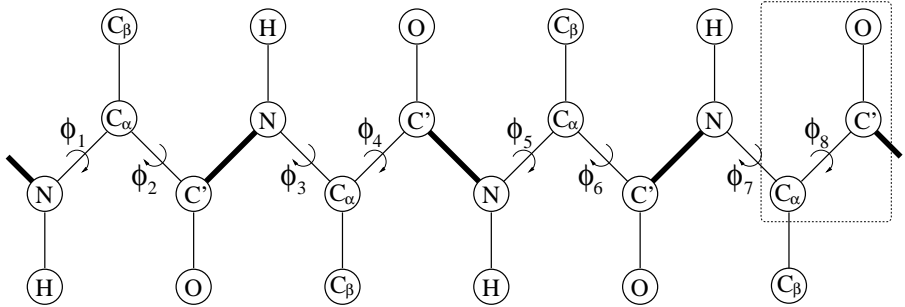}{12cm}{0}
\end{center}
\caption{Update of the chain defined in Sec.~\protect\ref{sec:2}.
Eight torsion angles $\phi_i$ are turned. Turns such that
the three atoms in the box are left unaffected are favored. Thick
lines represent peptide bonds and the peptide torsion angles   
are fixed.}     
\label{fig:1}\end{figure}

Consider a segment of four adjacent amino acids $k$, $k+1$, $k+2$ and
$k+3$ along the chain, and let the corresponding eight Ramachandran angles
(see Fig.~\ref{fig:1}) form a vector $\pv=(\phi_1,\ldots,\phi_n)$, where
$n=8$. A change $\dpv$ of $\pv$ will, by construction, leave all amino 
acids $k'<k$, as well as the N, H and \Ca\ atoms of amino acid $k$, 
unaffected. 
For all amino acids $k'>k+3$ to remain unaffected too, it is sufficient 
to require that the three atoms \Ca, \Cp\ and O of amino acid $k+3$ 
(see Fig.~\ref{fig:1}) do not move. If this condition is fulfilled,  
the deformation of the chain is local.   

Denote the position vectors of the \Ca, \Cp\ and O atoms of amino acid 
$k+3$ by $\rv_I$, $I=1,2,3$. A bias toward local deformations can 
be obtained by favoring changes $\dpv$ that correspond 
to small values of the quantity 
\beq
\Delta^2=\sum_{I=1}^3(\delta\rv_I)^2\,,
\label{delta2}\eeq
which for small $\delta\phi_i$ can be written as         
\beq
\Delta^2\approx\dpv^T\Gv\dpv=
\sum_{i,j=1}^n\delta\phi_iG_{ij}\delta\phi_j\,,
\eeq
where
\beq
G_{ij}=\sum_{I=1}^3\frac{\partial\rv_I}{\partial\phi_i}\cdot
\frac{\partial\rv_I}{\partial\phi_j}\,.
\label{Gij}\eeq
Note that the three vectors $\rv_I$ can be described in terms 
of six independent parameters, since bond lengths and angles are 
fixed. This implies that the $n\times n$ matrix $\Gv$, which 
by construction is non-negative and symmetric, has eigenvectors
with eigenvalue zero for $n=8>6$. A bias toward small 
$\Delta^2$ means that these soft modes are favored.   

We can now define the update, which consists 
of the following two steps. 
\begin{enumerate}
\item Draw a tentative new $\pv$, $\pv'$, from the Gaussian distribution 
\beq
W(\pv\to\pv')=\frac{(\det{\Av})^{1/2}}{\pi^3}
\exp\left[-(\pv'-\pv)^T\Av(\pv'-\pv)\right]\,,
\label{W}\eeq
where the matrix 
\beq
\Av=\frac{a}{2}(\unit+b\Gv)
\eeq
is a linear combination of the $n\times n$ unit matrix $\unit$
and the matrix $\Gv$ defined by Eq.~\ref{Gij}. 
The shape of this distribution depends on the parameters 
$a>0$ and $b\ge0$. The parameter $b$ sets the degree of
bias toward small $\Delta^2$. The bias is strong for large $b$
and disappears in the limit $b\to0$. The parameter $a$ is
a direction-independent scale factor that is needed to control the
acceptance rate. Larger $a$ means higher acceptance rate, for fixed $b$. 
If $b=0$, then the 
components $\delta\phi_i$ are independent Gaussian random numbers
with zero mean and variance $a^{-1}$.  
Note that $W(\pv\to\pv')\ne W(\pv'\to\pv)$ since the matrix $\Gv$ 
is conformation dependent.

\item Accept/reject $\pv'$ with probability 
\beq
\Pacc=\min\left(1,
\frac{W(\pv'\to\pv)}{W(\pv\to\pv')}\exp[-(E'-E)/kT]
\right)
\label{Pacc}\eeq
for acceptance. 
The factor $W(\pv'\to\pv)/W(\pv\to\pv')$ is needed for 
detailed balance to be fulfilled, since $W$ is asymmetric. 

\end{enumerate}

It should be stressed that this scheme is quite flexible.  
For example, it can be immediately applied to chains with
nonplanar peptide torsion angles. The use of the concerted-rotation 
method for simulations of such chains has recently been 
discussed~\cite{Dinner:00}. 

A convenient and efficient implementation of the algorithm can be 
obtained if one takes the ``square root'' of the matrix $\Av$, which
can be done because $\Av$ is symmetric and positive definite. More precisely,
it is possible to find a lower triangular matrix $\Lv$ (with nonzero
elements only on the diagonal and below) such that
\beq
\Av=\Lv\Lv^T\,.      
\label{chol}\eeq  
An efficient routine for this so-called Cholesky decomposition 
can be found in \cite{Press:92}.  

\subsection{Implementing  step 1}

Given the Cholesky decomposition of the matrix $\Av$, 
the first step of the algorithm can be implemented as follows.

\begin{itemize}

\item Draw a $\psv=(\psi_1,\ldots,\psi_n)$ from the distribution
$P(\psv)\propto\exp(-\psv^T\psv)$. The components $\psi_i$ are
independent Gaussian random numbers and can be generated, for example,
by using the Box-Muller method 
\beq 
\psi_i=(-\ln R_1)^{1/2}\cos2\pi R_2\,,    
\label{box}\eeq
where $R_1$ and $R_2$ are uniformly distributed random numbers
between 0 and 1. 

\item Given $\psv$, solve the triangular system of equations
\beq
\Lv^T\dpv=\psv
\label{lineq}\eeq
for $\dpv$. It can be readily verified that the $\dpv=\pv'-\pv$ 
obtained this way has the desired distribution Eq.~\ref{W}.

\end{itemize}

\subsection{Implementing step 2}

The Cholesky decomposition is also useful when calculating
the acceptance probability in the second step of the algorithm. 
The factor $W(\pv\to\pv')$ can be easily computed by using that 
\beq
(\det\Av)^{1/2}=\prod_{i=1}^n L_{ii}   
\label{det}\eeq
and that $\exp[-(\pv'-\pv)^T\Av(\pv'-\pv)]=\exp(-\psv^T\psv)$.
The reverse probability $W(\pv'\to\pv)$ depends on $\Av(\pv')$ and
can be obtained in a similar way, if one makes a Cholesky 
decomposition of that matrix, too. 

\subsection{Pivot update}

Previous simulations~\cite{Irback:00} of the model 
protein defined in Sec.~\ref{sec:2} were carried out by using 
simulated tempering with pivot moves as the elementary 
conformation update. With this algorithm, 
the system was successfully studied down 
to temperatures just below the folding transition. However, the 
performance of the pivot update, where a single angle $\phi_i$
is turned, deteriorates in the folded phase. What we hope is that 
the exploration of this phase can be made more efficient by alternating 
the pivot moves with moves of the type described previously.    

\section{Results}

The character of the proposed update depends strongly on the
bias parameter $b$. The suggested steps have a random direction  
if $b=0$. The distribution $W(\pv\to\pv')$ in Eq.~\ref{W} is,  
by contrast, highly asymmetric in the limit $b\to\infty$, with 
nonzero width only in directions corresponding to eigenvalue 
zero of the matrix $\Gv$. In particular, this implies that
the reverse probability $W(\pv'\to\pv)$ in the acceptance
criterion Eq.~\ref{Pacc} tends to be small for large $b$.

For the acceptance rate to be reasonable,  
it is necessary to use a very small step size if  
$b$ is small or large. The question is whether the step size can be 
increased by a better choice of $b$. To find that out, we performed 
a set of simulations of the three-helix-bundle protein defined in 
Sec.~\ref{sec:2} for different $a$ and $b$. Two different temperatures were 
studied, $kT=0.6$ and 0.7, one on either side of the folding temperature 
$kT_f\approx0.66$~\cite{Irback:00}.    

In these runs, we monitored the step size $S$, where 
\beq
S=|\dpv|=\left[\sum_{i=1}^n(\delta\phi_i)^2\right]^{1/2}
\label{size}\eeq
for accepted moves and $S=0$ for rejected ones. Measurements 
were taken only when the $n=8$ angles all were in 
the segment that makes the middle helix of the three. 
We focus on this segment because it is the most demanding part to update.  
        
The average step size, $\ev{S}$, depends strongly on $b$. 
A rough optimization of $b$ was carried out by maximizing $\ev{S}$ as
a function of $a$ for different fixed $b=10^k$ ($k$ integer).  
The best values found were $b_{\max}=10\ ({\rm rad/\AA})^2$ 
and $b_{\max}=0.1\ ({\rm rad/\AA})^2$ at 
$kT=0.6$ and $kT=0.7$, respectively. Note that the preferred 
degree of bias is higher in the folded phase. 
  
\begin{figure}[t]
\hspace{0mm}
\mbox{
  \fig{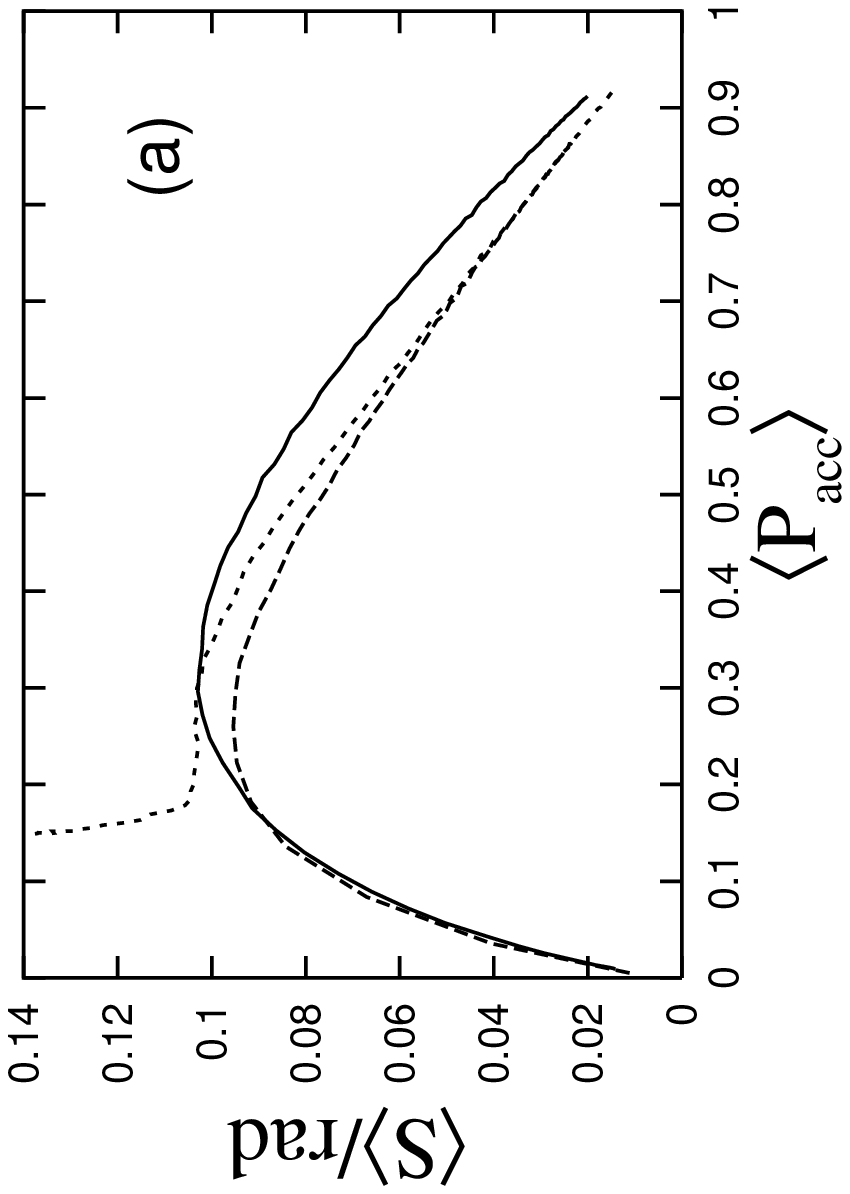}{7cm}{-90}
  \fig{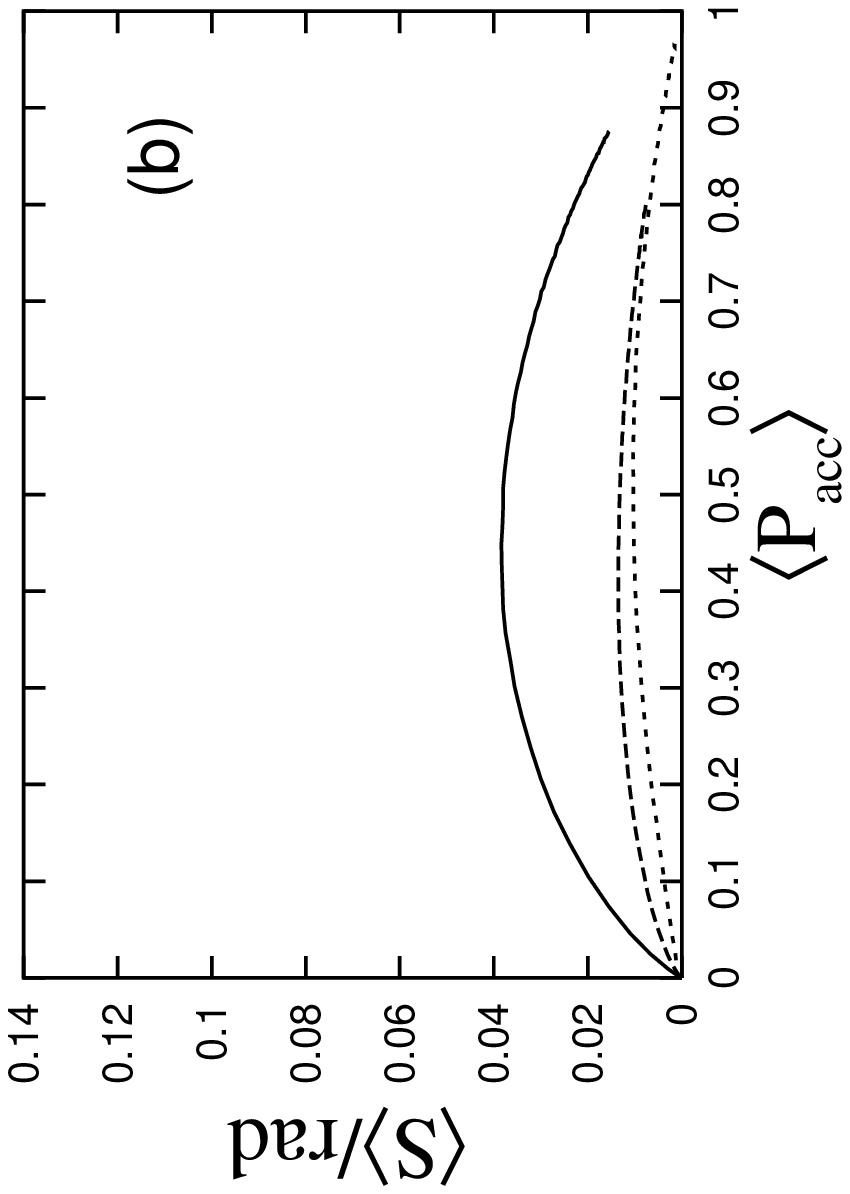}{7cm}{-90}
}
\caption{Average step size, $\ev{S}$, against average acceptance 
rate, $\ev{\Pacc}$, for different updates at (a) $kT=0.7$ and 
(b) $kT=0.6$. Shown are results for the $b=b_{\max}$ (full lines),
$b=0$ (dashed lines) and pivot (dotted lines) updates.}  
\label{fig:2}\end{figure}

In Fig.~\ref{fig:2}, we show $\ev{S}$ against the average
acceptance rate, $\ev{\Pacc}$, for $b=0$ and $b=b_{\max}$ 
at the two temperatures;
$\ev{\Pacc}$ is an increasing function of $a$ for fixed $b$ and $T$.  
Also shown are the corresponding  
results for the pivot update, where only one angle $\phi_i$ is turned 
($S=|\delta\phi_i|$ if the change is accepted). 
At the higher temperature, we find that the
$b=b_{\max}$ and $b=0$ updates show similar behaviors.
The pivot update is somewhat better and has its maximum 
$\ev{S}$ at low $\ev{\Pacc}$, where the proposed change $\delta\phi_i$ 
is drawn from the uniform distribution between 0 and $2\pi$.
This is consistent with the finding~\cite{Madras:88} that
the pivot update is a very efficient method for self-avoiding walks,  
in spite of a low acceptance rate. 
The situation is different at the lower temperature, which is much 
harder to simulate. Here, the $b=b_{\max}$ update is the best. 
The maximum $\ev{S}$ is approximately three times higher for this
method than for the other two. This shows that the biasing probability 
Eq.~\ref{W} is indeed useful in the folded phase.   

The $b=0$ update can be compared with the moves used by 
Shimada~\etal~\cite{Shimada:00} in a recent all-atom study of 
kinetics and thermodynamics for the protein crambin with 46 amino acids.
These authors updated sets of two, four or six backbone torsion angles,
using independent Gaussian steps with a standard deviation of $2^\circ$. 
Our $b=0$ update has maximum $\ev{S}$ at $a\approx 6400\ ({\rm rad})^{-2}$
for $kT=0.6$, which corresponds to a standard 
deviation of $0.7^\circ$.  This value is in line with that used by 
Shimada~\etal, since we turn eight angles.   

\begin{figure}[t]
\hspace{35mm}
  \fig{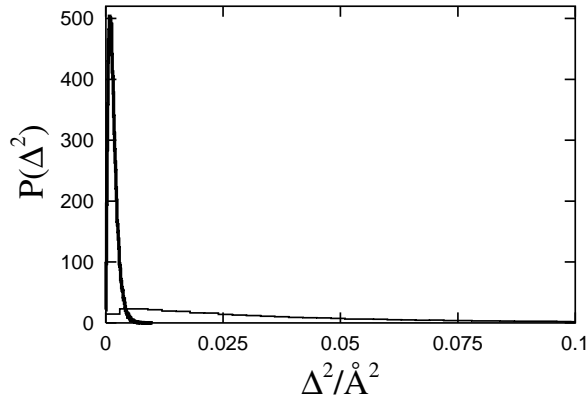}{8cm}{-90}
\caption{Distributions of $\Delta^2$ (see Eq.~\protect\ref{delta2}) 
for the $b=b_{\max}$ (thick line) and $b=0$ (thin line) updates at $kT=0.6$.
The values used for the parameter $a$ correspond to maximum $\ev{S}$.}
\label{fig:3}\end{figure}

How local is the method for $b=b_{\max}$?
To get an idea of that, we calculated the distribution of $\Delta^2$ 
(see Eq.~\ref{delta2}) for accepted moves, for $b=b_{\max}$ and $b=0$ at
$kT=0.6$. As was previously the case, we restricted ourselves to angles 
in the middle helix. The two distributions are shown in Fig.~\ref{fig:3}  
and we see that the one corresponding to $b=b_{\max}$ 
is sharply peaked near $\Delta^2=0$. This shows
that the $b=b_{\max}$ update is much more local than the 
unbiased $b=0$ update, although the average step size, 
$\ev{S}$, is considerably larger for $b=b_{\max}$.

\begin{figure}[t]
\hspace{35mm}
  \fig{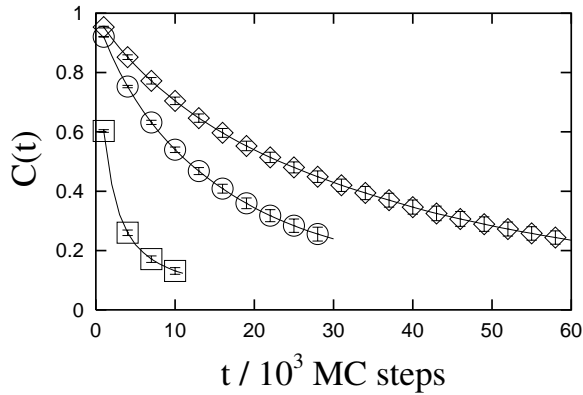}{8cm}{-90}
\caption{The autocorrelation function $C(t)$ (see the text) at $kT=0.6$ 
for the $b=b_{\max}$ ($\Box$), $b=0$ ($\circ$) and pivot ($\Diamond$) 
updates. Step-size parameters correspond to maximum $\ev{S}$.}  
\label{fig:4}\end{figure}

So far, we have discussed static (one-step) properties of the
updates. We also estimated the dynamic autocorrelation function
\beq
C_i(t)=\frac{\ev{\cos\phi_i(t)\cos\phi_i(0)}-\ev{\cos\phi_i(0)}^2}
{\ev{\cos^2\phi_i(0)}-\ev{\cos\phi_i(0)}^2} 
\eeq  
for different $\phi_i$. This measurement is statistically very difficult  
at low temperatures. However, the sixteen most central angles 
$\phi_i$ in the sequence, all belonging to the middle helix, were found 
to be effectively frozen at $kT=0.6$, and the time scale for the 
small fluctuations of these angles about their mean values was 
possible to estimate. In Fig.~\ref{fig:4}, we show the average
$C_i(t)$ for these sixteen angles, denoted by $C(t)$, against        
Monte Carlo time $t$, for the $b=0$, $b=b_{\max}$ and pivot updates. 
One time unit corresponds to one elementary move, accepted or rejected,
at a random position along the chain. 
We see that $C(t)$ decays most rapidly for the $b=b_{\max}$ 
update. So, the larger step size of this update does make 
the exploration of these degrees of freedom more efficient.   

Let us finally comment on our choice to work with $n=8$ angles.
This number can be easily altered and some calculations were done
with $n=6$ and $n=7$, too. For $n=6$, the performance was worse, 
which is not unexpected because there are no soft modes 
available; there are not more variables than constraints. 
The results obtained for $n=7$ were, by contrast, comparable 
to or slightly better than the $n=8$ results. 

\section{Discussion}
 
Straightforward Monte Carlo updates of torsional degrees of freedom
tend to cause large changes in the global structure of the chains
unless the step size is made very small, which is a problem 
in simulations of dense polymer systems. The strictly local 
concerted-rotation approach provides a solution to this 
problem but is rather complicated to implement. In this paper, 
we have discussed a method that may be less powerful 
but is much easier to implement, which suppresses
rather than eliminates nonlocal deformations. 
                                                            
The method is flexible and not much harder to implement than simple 
unbiased updates. However, compared to such updates, it has two 
distinct advantages: the step size can be increased and 
the update becomes more local, as shown by our simulations of
the three-helix-bundle protein in its folded phase. 

Making the update more local is important in order to be able 
to increase the step size and thereby improve the efficiency. 
At the same time, it makes the dynamics more realistic; the proposed 
method is, in contrast to the other methods mentioned, tailored to avoid 
drastic deformations both locally and globally. Therefore, although this 
paper was focused on thermodynamic simulations, it should be 
noted that this method may be useful for kinetic studies, too.

\subsection*{Acknowledgments}

This work was supported in part by the Swedish Foundation for Strategic 
Research. G.F. acknowledges support from Universit\'a
degli studi di Cagliari and the EU European Social Fund.

\end{document}